# Slowly cooling white dwarfs in M13 from stable hydrogen burning

Jianxing Chen[1,2], Francesco R. Ferraro[1,2*], Mario Cadelano[2], Maurizio Salaris[3], Barbara Lanzoni[1,2], Cristina Pallanca[1,2], Leandro G. Althaus[4,5], Emanuele Dalessandro[2]

[1] Dipartimento di Fisica e Astronomia "Augusto Righi", Alma Mater Studiorum Universita` di Bologna, via Piero Gobetti 93/2, I-40129 Bologna, Italy

[2] INAF -- Astrophysics and Space Science Observatory Bologna, Via Gobetti 93/3, I-40129 Bologna, Italy

[3] Astrophysics Research Institute, Liverpool John Moores University, Liverpool Science Park, IC2 Building, 146 Brownlow Hill, Liverpool L3 5RF, UK

[4] Grupo de Evolucion Estelar y Pulsaciones, Facultad de Ciencias Astronomicas y Geofisicas, Universidad Nacional de La Plata, Paseo del Bosque s/n, 1900 La Plata, Argentina

[5] CCT – CONICET Centro Cientıfico Tecnologico La Plata, Consejo Nacional de Investigaciones Cientıficas y Tecnicas, Calle 8 No. 1467, B1904CMC La Plata, Buenos Aires, Argentina

**White Dwarfs (WDs) are the final evolutionary product of the vast majority of stars in the Universe. They are electron-degenerate structures characterized by no stable thermonuclear activity, and their evolution is generally described as a pure cooling process. Their cooling rate is adopted as cosmic chronometer to constrain the age of several Galactic populations, including the disk, globular and open clusters. By analysing high-resolution photometric data of two twin Galactic globular clusters (M3 and M13), we find a clear-cut and unexpected over-abundance of bright WDs in M13. Theoretical models suggest that, consistently with the horizontal branch morphology, this over-abundance is due to a slowing down of the cooling process in ~70% of the WDs in M13, caused by stable thermonuclear burning in their residual hydrogen-rich envelope. This is the first observational evidence of quiescent thermonuclear activity occurring in cooling WDs and it brings new attention on the use of the WD cooling rate as cosmic chronometer for low metallicity environments.**

White Dwarfs (WDs) are the electron-degenerate core remnants of low-mass stars that have concluded their thermonuclear activity: all stars below 8 $M_\odot$, with a possible extension up to 11 $M_\odot$, are expected to end their evolution as WDs[1,2]. Their study provides a large amount of information on the physical properties and the evolution of their progenitors, which constitute ~98% of all the stars in the Universe. They are thought to be characterized by the absence of thermonuclear activity and, in fact, in the vast majority of theoretical models[3,4] stable nuclear burning in the residual hydrogen-rich envelope (with mass thickness in the range $10^{-4} - 10^{-7}$ $M_{WD}$) is assumed to be negligible[5,6] and their evolution, in essentially all the astrophysical textbooks, is described as a pure cooling process. Because of this relatively simple structure and evolution, the faint-end of the WD cooling sequence has been proposed as accurate cosmic chronometer to constrain the age of several Galactic populations, including the disk, globular and open clusters. For this reason, a notable observational effort with the HST has been devoted to the exploration of the WD cooling sequence in Galactic stellar systems[7-12], especially in globular clusters (GCs) that are the oldest and richest ones. This is because, in spite of their relatively large distances (typically ~ 10 kpc), GCs provide large and homogeneous samples of WDs from coeval progenitors, all located at the same distance from the observer[13].

However, recent computations[14-16] demonstrate that even a relatively small amount of residual hydrogen (a few $10^{-4}$ $M_\odot$) left by the previous evolutionary stages is sufficient to allow quiescent thermonuclear burning. For low-mass (<0.6 $M_\odot$) and low-metallicity (Z<0.001) WDs this can provide a non-negligible source of energy (larger than 40%) in the brightest portion of the cooling sequence, which then rapidly becomes negligible at $\log(L/L_\odot) \approx -4$ (log Te ≈ 3.7; see Extended Data Figure 1). This (usually neglected) energy source slows down the cooling process, with a consequent observable impact on the WD luminosity function (LF), since an increased cooling time naturally translates into a larger number of WDs. The delay in cooling time cumulates during the H-burning phase and reaches a value as large as ~ 760 Myr, which then remains constant and affects the reading of the cooling time at fixed luminosity along the entire cooling sequence (see Extended Data Figure 1). Models[14-16] show that the phenomenon is particularly relevant in the low metallicity regime (Z<0.001). Hence it should poorly affect the high-metallicity WDs orbiting the Galactic disk, while it is expected to become observable in the typical metallicity regime of old GCs populating the Galactic halo.

**Results -** With the aim of constraining the physics of hot WDs and their cooling processes, here we present the detailed comparison between the brightest portion of the WD cooling sequence in two twin old and massive GCs: M3 and M13. These systems represent a classical "horizontal branch (HB) morphology pair"[17,18], because they share many physical properties (see Extended Data Fig.2), as the metallicity ([Fe/H]~ −1.6) and the age (t ~ 13 Gyr)[19-22], but, at the same time, they display quite different HB morphologies: a very pronounced blue tail is present in the colour-magnitude diagram (CMD) of M13, while it is absent in M3. The origin of such a difference is not completely understood yet, although various hypotheses have been suggested[23,24]. For the present study, we used a set of ultra-deep HST observations of these two clusters, acquired in the near-ultraviolet with the UVIS channel of the Wide Field Camera 3 (WFC3; see Section 'Data-set and Analysis' in Methods). The resulting ($m_{F275W}$, $m_{F275W}$ − $m_{F336W}$) CMDs of M3 and M13 are shown in Figure 1. Their superb quality provides a full view of the stellar populations of the two systems down to $m_{F275W}$ = 25, allowing a clear definition of all the evolutionary sequences. As expected in old stellar populations, the bluest portion of the HB dominates in the F275W band, blue stragglers are comparable in magnitude to red giant branch (RGB) stars, and bright WDs reach approximately the same luminosity of main sequence turn-off (MS-TO) stars (see Section 'M3 and M13: differences and similarities' in Methods). Indeed, a quite well defined WD cooling sequence, extending by 5 magnitudes in the F275W filter, is clearly visible in the blue side of both CMDs. To make the comparison between the WD cooling sequences of the two clusters as straightforward as possible, we shifted the CMD of M3 to match that of M13. We found that only a shift in magnitude ($\Delta m_{F275W}$ = −0.55 ± 0.02) is required (see Section 'M3 and M13: differences and similarities' in Methods), providing an impressive match of all the evolutionary sequences and further confirming that the two systems have approximately (within a few 0.1 Gyr) the same age. At this point the two CMDs are fully homogeneous and we can focus on the properties of the WD cooling sequences: the two CMDs (after alignment) zoomed on the WD region are shown in Figure 2. In the following analysis, we conservatively selected the WD samples down to a magnitude limit $m_{F275W}$ ≤ 23.5, where the completeness level is larger than 50% at any distance from the centre in both clusters (see Section 'Artificial star experiments and completeness' in Methods and Extended Data Fig.3), and we excluded the stars located at more than 3σ from the mean ridge line of the cooling sequence (σ being the photometric error at each magnitude level). The comparison with theoretical WD cooling sequences[3] suggests that the adopted magnitude cut corresponds to a luminosity $\log(L/L_\odot)$ ~ −1.7, which limits the following analysis to the first ≈ 100 Myr of cooling (see Fig. 2).

The adopted selection criteria provide 418 and 284 WDs in M13 and M3, respectively. Their LFs, both before and after correction for incompleteness, and computed in bins of 0.5 magnitudes are shown in Figure 3a. It can be seen that the conservative assumptions adopted for the sample selection strongly limit the impact of incompleteness: the global correction to the adopted samples is smaller than of 15%, with completeness-corrected populations counting 467 and 326 WDs in M13 and M3, respectively. Hence, the completeness correction does not significantly alter the ratio between the number of WDs in the two clusters, with M13 showing a population that is approximately 1.4 times larger than that of M3. Furthermore, also the luminosity distribution along the cooling sequences appears to be different. To quantify this, we built the normalized $m_{F275W}$ cumulative distributions of the two samples (Figure 3b) and performed a Kolmogorov-Smirnov test. We found that the probability that the two WD samples are extracted from the same parent population is $9 \times 10^{-3}$, indicating that the discrepancy is indeed significant and worth to be investigated in detail.

To take into account the different intrinsic richness of the two clusters, we normalized the observed samples of WDs to the RGB population (see Section "Normalizing the WD samples' in Methods and Extended Data Fig.4), finding that the global population ratios are $N_{WD}/N_{RGB} = 467/1176 = 0.40 \pm 0.02$ in M13 and $N_{WD}/N_{RGB} = 326/1235 = 0.26 \pm 0.02$ in M3. This corresponds to a factor $\sim 1.5$ more WDs per RGB star (or unit mass) in the former. Further insights can be obtained from the WD LFs normalized to the RGB reference populations (namely, the number of WD counted in each magnitude bin, divided by the total number of RGB stars in the same cluster), which is shown in Figure 4. This comparison clearly confirms the result above.

To understand the origin of this difference we first need to assess whether it is due to an excess of WDs in M13, or a deficit of WDs in M3. We thus compared the observations to theoretical expectations. The selected portion of the cooling sequence is made of WDs formed at most $\sim 80$ Myr ago (see Figure 2), while the selected portion of the RGB population samples a time interval[25] of $\sim 320$ Myr (see Section "Normalizing the WD samples' in Methods). Given that the ratio between the number of WDs and RGB stars is proportional to the ratio of the corresponding lifetimes, we should find $N_{WD}/N_{RGB} = t_{WD}/t_{RGB} = 80/320 = 0.25$, largely independent of the WD mass and the details of the WD structure (see Section 'WD lifetime ratio' in Methods). The number ratio in M3 ($N_{WD}/N_{RGB} = 0.26 \pm 0.02$) is fully consistent with the theoretical expectation[3,25], whilst the value measured in M13 is significantly larger, $N_{WD}/N_{RGB} = 0.40 \pm 0.02$. The comparison of the normalized differential LFs and the corresponding theoretical expectation based on the evolutionary time ratio of WDs[3] and RGB stars[25] (black dashed line in Figure 4) further confirms that the discrepancy pertains to the faintest portion of surveyed cooling sequence of M13. We can thus conclude that the detected difference between the two clusters is due to a significant excess of WDs in M13 (see also Section 'Excess of WDs or lack of RGBs?' in Methods).

**Discussion**

As first step, we can safely exclude that the detected WD excess in M13 arises from the long-term internal dynamical evolution of the cluster (as suggested[8], instead, in the case of NGC 6397), because the sampled cooling time (80 Myr) is much smaller than the cluster central relaxation time ($t_{rc}$~320 Myr; see Section 'The different BSS and HB content in M3 and M13' in Methods and Extended Data Figure 2).

An increase in the number of WDs could be due to a slower cooling process, or an additional channel of WD formation. The mechanism needs to be active just in M13 and could therefore be related to other well-known differences between these two "twin" systems. Indeed, the observed difference in the WD populations could be linked to the different morphologies of the HB. In fact, because of their small envelope mass, the bluest HB stars are expected to completely or partially skip the subsequent asymptotic giant branch (AGB) phase, with a significant impact on the size of the residual hydrogen-envelope in the forming WD and its cooling time[16].

The adopted BaSTI theoretical models[25] show that, at metallicity Z=0.001, HB stars with masses smaller than ~ 0.56 $M_\odot$ do not experience the thermal pulse-AGB stage, where helium and hydrogen are alternatively burned in two separate shells surrounding the carbon-oxygen core, and during which the third dredge-up can occur. This is the deepest convective event occurring in a star's life. It takes place when the (more external) hydrogen-burning shell is temporarily turned off and the surface convection penetrates in the stellar interior extending down to the inter-shell region, which was previously mixed by inner convection during a temporary ignition of the He-burning shell (a thermal pulse). During the third dredge-up, a significant amount of carbon is carried into the convective envelope and, in turn, a significant amount of hydrogen is brought down inside the star, and there it is burned. Thus, the occurrence (or not) of the third dredge-up affects the residual mass of hydrogen with which the proto-WD reaches its cooling track, consequently impacting the WD structure and cooling time. More specifically, appropriate theoretical models[16] found that the proto-WD stars skipping the third dredge-up have hydrogen envelopes thick enough to guarantee stable hydrogen-burning during the WD cooling evolution, with a resultant extra-energy production that increases the cooling time. At the metallicity of M3 and M13, the minimum thickness needed[14] to sustain hydrogen-burning is ~$1.7 \times 10^{-4}\,M_\odot$ and the cooling time is increased by ~75% at the minimum WD luminosity sampled in our study ($\log(L/L_\odot) \sim -1.7$; see Extended Data Fig. 5).

These arguments support the fact that, along with "standard WDs" well described by models[3,4] with no hydrogen-burning, also "slow WDs" may exist, generated by progenitors that do not experience the third dredge-up during the thermal pulses. Due to their longer cooling times, these slow WDs could explain the observed pile-up along the WD LF in M13. The HB morphology of this cluster and its AGB-HB population ratio (see Section 'AGB manqué stars in M13' in Methods and Extended Data Fig. 6) seem to suggest that a fraction of its stars does indeed skip this evolutionary stage and could therefore give origin to slow WDs. In fact, not only some AGB manqué star (i.e., stars missing the AGB phase completely) is observed in the CMD (see Section 'AGB manqué stars in M13' in Methods), but also other objects are expected to leave the AGB before the onset of thermal pulses. Overall, theoretical models predict that all the stars with HB mass smaller than 0.56 $M_\odot$ (this limit depending on the metallicity) do not experience the thermal pulses (hence the third dredge-up) and can thus generate slow WDs. To estimate the fraction of HB stars with mass below this threshold in M13 and M3 we adopted the HB mass distributions determined in previous studies[24] that not only match the observed number distribution of HB stars in magnitude and colour, but also reproduce, within 1%, the observed $N_{AGB}/N_{HB}$ ratio of both clusters. We found that only a very small percentage (well below 10%) of HB stars has masses below 0.56 $M_\odot$ in M3, thus suggesting that the WD population in this cluster descends from objects that experienced the third dredge-up. The fraction, instead, increases up to 65% in M13.

Putting all these arguments together, we should expect that in M13 (and possibly in any cluster with extended blue HB), two populations of WDs exist:

(1) standard WDs, whose progenitors experienced thermal pulses and third dredge-up, reaching the cooling sequence with a hydrogen envelope not massive enough to ignite stable burning;

(2) slow WDs, originated by stars with HB mass smaller than ~ 0.56 $M_\odot$ that skipped the third dredge-up and maintained a residual hydrogen envelope thick enough to allow burning during the subsequent WD evolution, significantly increasing their cooling times.

Could this double WD formation channel be at the origin of the bright WD excess detected in M13? To quantitatively test this scenario we have performed Monte Carlo simulations of the entire RGB-HB-WD evolutionary path appropriate for M3 and M13 (see Section 'Monte Carlo simulation' in Methods), randomly extracting a large set (one million) of age values from a uniform distribution between zero (the base of the adopted RGB selection box; see Extended Data Fig. 4) and the maximum WD age at the bottom of the considered cooling sequence ($m_{F275W}$ = 23.5), and determining the position of the corresponding synthetic stars along the full (RGB-HB-WD) evolutionary sequence.

For the WDs in M3 we used only cooling models[16] with no hydrogen burning, and we found $N_{HB}/N_{RGB}$ = 0.29 and $N_{WD}/N_{RGB}$ = 0.25, in very good agreement with the observed values (0.28 ± 0.02 and 0.26 ± 0.02, respectively). This fully confirms that essentially all the WD progenitors in M3 experienced the third dredge-up. In M13 we imposed that 35% of the synthetic HB stars generate standard WDs and 65% give origin to slow WDs, described, respectively, by cooling models[16] with non-active and active hydrogen burning. Given the longer cooling times of the WDs from the "slow channel", we find that they contribute to 70% of the total WD population in the relevant magnitude range. From these simulations we obtained $N_{HB}/N_{RGB}$ = 0.34, in satisfactory agreement with the observed value (0.31 ± 0.02), and $N_{WD}/N_{RGB}$=0.37 that now compares well with the observed ratio (0.40 ± 0.02). Moreover, the observed cumulative LF in M13 is impressively well matched by the simulations including a majority of slow WDs, while it turns out to be completely inconsistent with the curve obtained by assuming 100% of standard WDs (see Figure 5).

**Conclusions and future perspectives** – Our results suggest that residual hydrogen-burning on the WD surface is the physical phenomenon responsible for the bright WD excess observed in M13. The fact that this phenomenon occurs in M13 and not in M3 is well consistent with the different HB morphology of the two clusters. Actually, at odds with M3, the HB of M13 shows a long extension to the blue that is populated by lower mass stars; these are destined to skip the third dredge-up occurring along the AGB, during which most of the hydrogen remaining in the stellar envelope is consumed. Thus, a significant fraction of the stars in M13 are expected to end their evolution as WDs characterized by a hydrogen envelope massive enough to allow thermonuclear burning that, in turn, slows down the cooling process and consequently affects the observed LF. At the moment, this appears to be the most viable and natural explanation, while alternative scenarios should invoke ad hoc and unknown mechanisms able to increase the production or slow down the cooling process of the WDs in M13, and not in M3. The discovery reported in this paper represents the first direct evidence for the occurrence of stable nuclear burning in the residual hydrogen envelope of cooling WDs and offers an empirical measure of the delay in the flow of time marked by the WD clock in the presence of slowly cooling WDs. Even considering the relatively small portion of the cooling sequence studied here, the effect turns out to be notable: the cooling age at $m_{F275W}$=23.5, (corresponding to $\log(L/L_\odot) \sim -1.7$) as read from canonical WD models is 80 Myr and increases to 140 Myr (Δt=60 Myr, corresponding to an increment of 75%) when WD models with active hydrogen-burning are adopted.

Systematic and quantitative studies (even of the bright portion, only) of the WD cooling sequence in star clusters with different HB morphologies and different metallicities are now urged to firmly establish and empirically characterize the connection between the HB morphology and the WD excess, and its dependence on the metal content of the system. This will provide the needed observational constraints to advanced models of WD evolution, and it will allow a correct use of WDs as cosmic chronometers, removing systematic uncertainties that can be as large as 1 Gyr.

# Methods

**Data-set and Analysis:** For this study we have used a set of ultra-deep, high-resolution images acquired with the UVIS channel of the Wide Field Camera 3 (WFC3) on board the Hubble Space

Telescope. The data-set in each cluster is composed of 6 images (each one with exposure time of 415 s in M3, and 427 s in M13) in the F275W filter, and 4 images (each one 350 s long in both clusters) in the F336W.

The photometric analysis was performed via the point-spread function (PSF) fitting method, by using DAOPHOT IV[27] and following the "UV-route" approach used in previous works[28-34]. Briefly, the photometric analysis was carried out on the "_flc" images (which are the UVIS calibrated exposures, including Charge Transfer Efficiency correction) and it consists in first searching for the stellar sources in the near-UV images, then force-fit the source detection at longer wavelengths at the same positions of the UV-selected stars. Because the crowding effect generated by giant and turn-off stars (which are increasingly brighter at increasing wavelengths) is strongly mitigated in the near-UV images of old stellar populations (as Galactic globular clusters), this procedure allows the optimal recovering of blue and faint objects, like WDs. So, at first, we selected 200-300 bright and unsaturated stars, relatively uniformly distributed in the sampled field of view, to determine the PSF function of each F275W exposure. The resulting model was then applied to all the sources detected above $5\sigma$ from the background, and the stars found at least in half of the UV images were combined to create a master list. The photometric fit was then forced in all the other frames at the positions corresponding to the master list stars, by using DAOPHOT/ALLFRAME[35]. Finally, the magnitudes estimated for each star in different images of the same filter were homogenized, and their weighted mean and standard deviation have been adopted as the star magnitude and its related photometric error. The instrumental magnitudes were calibrated to the VEGAMAG system and the instrumental coordinates have been put onto the International Celestial Reference System by cross-correlation with the catalogue obtained from the HST UV Globular Cluster Survey[36]. The analysis of the sharpness quality parameter of the PSF fitting procedure shows no significant evidence of non-stellar sources in the WD sample under analysis ($m_{F275W}<23.5$).

**M3 and M13: differences and similarities -** The CMDs shown in Figure 1 clearly illustrates the well-known similarities and differences between the two clusters. First of all, the different HB morphology, with a lack of stars redder than ($m_{F275W} - m_{F336W}$) = 0.5 and a blue tail extension becoming apparent at colours bluer than ($m_{F275W} - m_{F336W}$) = 0 in M13. A copious population of RR Lyrae (observed at random phases) is, instead, clearly visible in the CMD of M3, as a sort of stream of stars at $m_{F275W} > 16.5$ and ($m_{F275W} - m_{F336W}$) > 0.8. Also, the different blue straggler star (BSS) content is well apparent, with a significantly larger population in M3 compared to M13[17,37-39].

Apart from these differences, the overall morphology of the CMDs appears to be very similar. To make the comparison more direct, we shifted the CMD of M3 to match that of M13. Indeed the two clusters share many physical properties (see Extended Data Figure 1) and, in particular, they have approximately the same metallicity ([Fe/H]∼ −1.6) and age (t ∼ 13 Gyr)[19-22]. Hence, the MS-TO and the subgiant branch appear to be optimal reference sequences in the UV-CMD to determine the shifts in magnitude and colour needed to make the two CMDs coincident. We found that only a shift in magnitude ($\Delta m_{F275W}$ = −0.55 ± 0.02) is required to superpose the CMD of M3 onto that of M13, providing an impressive match of all evolutionary sequences.

**Artificial star experiments and completeness.** For a quantitative study of the WD populations in terms of star counts, it is necessary to take into account the completeness of the photometric catalogues. We thus evaluated the completeness level of the stars observed along the WD cooling sequences in M3 and M13, at different magnitudes and different distances from the cluster centres, by performing artificial stars experiments. To this end, we followed the standard prescriptions[40-42] that we briefly summarize here. For each cluster, we created a list of artificial stars with a F275W (input) magnitude extracted from a LF modelled to reproduce the observed one. Then, each of these stars was assigned a F336W magnitude obtained by interpolating along the mean ridge line of the WD cooling sequence in the ($m_{F275W}$, $m_{F275W} - m_{F336W}$) CMD. The artificial stars thus generated were then added to the real images by using the DAOPHOT/ADDSTAR software, and the entire photometric analysis was repeated following exactly the same steps described in Section 'Data-set and Analysis' above. To avoid artificial crowding, the added stars were placed into the frames in a regular grid of 23×23 pixels (corresponding to about fifteen times the FWHM of the stellar sources), each cell containing only one artificial star during each run. The procedure was iterated several times and more than 150,000 artificial stars have been overall simulated in the entire field of view sampled by each cluster.

The ratio between the number of artificial stars recovered at the end of the photometric analysis ($N_o$) and the number of stars that were actually simulated ($N_i$) defines the completeness parameter $C = N_o/N_i$. As well known, the value of C strongly depends on both crowding and stellar luminosity, becoming increasingly smaller in cluster regions of larger density and at fainter magnitudes. We thus divided the sample of simulated stars in bins of cluster-centric distances (stepped by 5″) and F275W magnitudes (from ∼ 18, to ∼ 25, in steps of 0.5 mags) and, for each cell of this grid, we counted the number of input and recovered stars, calculating the corresponding value of C. The sizes of the radial and magnitude bins were chosen to secure enough statistics while keeping as high as possible both the spatial resolution and the sensitivity of the completeness curve to changes in the stellar luminosity. The uncertainties assigned to each completeness value ($\sigma_C$) were then computed by propagating the Poisson errors, and typically are on the order of 0.05. The construction of such a completeness grid allowed us to assign the appropriate C value to each observed WD with given F275W and F336W magnitudes, located at any distance from the cluster centre (see Extended Data Fig. 3). As expected, C progressively decreases toward the cluster centre and for fainter and fainter magnitudes. However, the completeness level remains relatively high down to very faint magnitudes, being larger than 50% for $m_{F275W} \sim 24$ and $m_{F275W} \sim 24.2$ in M3 and M13, respectively.

**Normalizing the WD samples** – To take into account the intrinsic richness of the two clusters, we used as normalization the number of stars counted in a post-main sequence stage, which is directly observable, independent of parameters like distance and reddening, expected to be the same in systems of equal total mass, age and metallicity, and only dependent on stellar evolutionary times[43]. In particular, because of their large luminosities (corresponding to large completeness levels at any distance from the centre), we adopted as reference the RGB population. Taking advantage of the excellent alignment obtained between the CMDs of the two clusters, we used the same selection box, sampling the entire RGB extension nearly down to base of the RGB (located at $m_{F275W} \sim 18.9$) up to the tip (see Extended Data Fig. 4). We counted 1176 RGB stars in M13 and 1235 in M3. According to the BaSTI evolutionary track[25] of a 0.8 $M_\odot$ star with appropriate metallicity, the time needed to evolve across the selected portion of the RGB is $\sim 320$ Myr.

**WD lifetime ratio** – The derived lifetime[3,25] ratio between the cluster WDs down to the considered limiting magnitude ($m_{F275W} \sim 23.5$) and the selected RGB stars is equal to $t_{WD}/t_{RGB} = 0.25$, independent of the adopted WD mass for values between 0.50 and 0.60 $M_\odot$, which are appropriate for the present study. The same ratio, and in general the same cooling times at each point across the relevant magnitude range, are found when we employ independent calculations by other authors[4], both their models with hydrogen mass fraction equal to $10^{-4}$ times the total WD mass (like in the adopted[16] computations), and those with much thinner hydrogen layers ($10^{-10} \times M_{WD}$).

**Excess of WDs or lack of RGBs?** In principle, the large $N_{WD}/N_{RGB}$ ratio measured in M13 does not necessarily imply an excess of WDs in this cluster: it could also be explained by a severe lack of RGB stars. Indeed, because of some mass-loss enhancement during this phase, a fraction of RGB stars in M13 could evolve directly to the helium-core WD stage before the helium flash: after all, the bluest objects along the cluster HB have masses only slightly larger than the helium-core mass at helium ignition, as discussed below. However, the study[44] of the luminosity function of RGB stars in M13 discloses no convincing evidence that a sizeable fraction of stars leaves the RGB before undergoing the helium flash. We can thus safely conclude that M13 hosts a genuinely larger population of WDs with respect to M3.

**The different BSS and HB content in M3 and M13** - It has been noted in many previous papers[17,37-39], that these two clusters host quite different populations of blue stragglers. This is apparent also in the CMDs presented in Figure 1, where we count 153 BSSs in M3 and only 62 in M13, if we adopt the same prescriptions and selection box used in previous works[38]. In addition, the BSS radial distribution, which can be used as "dynamical clock"[37-39], is different in the two systems, testifying that M3 is more dynamically evolved than M13, also in agreement with the values of the central relaxation time derived from the cluster structural parameters[45] (see Extended Data Figure 2). The relaxation times of both systems are larger than the age of the oldest WDs here considered, denoting that the observed difference cannot be due to internal dynamical processes. In general, it is seemingly hard to associate the different blue straggler properties and the different cluster dynamical stages to the observed excess of bright WDs in M13.

We have then investigated the role that the well-known difference in the HB morphology might play. As it is apparent in Figure 1, the HB of M13 has a long extension to the blue (hot temperatures), at odds with that of M3, and many studies in the literature have tried to explain this difference. A detailed comparison[23] between the two clusters, concluded that, independent of the adopted mass-loss recipe, the HB morphology cannot be reproduced in terms of an age difference[19-22], which (if any) must be very small. Hence, at least one additional parameter should be invoked to account for the blue HB extension observed in M13, and subpopulations of helium-enhanced stars are in principle a viable solution. In fact, for any given cluster age, MS-TO and RGB stars belonging to a helium-enhanced subpopulation have masses lower than their helium-normal counterparts, thus producing bluer HB stars (for fixed RGB mass-loss) with thinner envelopes around the helium cores. A recent detailed analysis[24] estimated that, to account for the observed HB morphologies, the HB stars in M3 had to have a spread $\Delta Y \sim 0.02$ in their initial helium mass fraction, whilst in the case of M13 $\Delta Y \sim 0.05$. A larger Y spread among the HB stars of M13 was estimated also in other similar analysis[22]. Other authors[46] instead, found that the maximum internal helium variation between the reddest and bluest HB stars in each of the two clusters is essentially the same, implying a difference of just 0.01 in the helium mass fraction of the two systems. Although these results demonstrate that the origin of the discrepancy it still not clear, as a matter of fact the HB morphology is strikingly different in the two clusters.

**AGB manqué stars in M13** - Interestingly, the extreme blue group in the HB blue-tail of M13 defines a sort of vertical sequence (here named extreme-HB) that counts ∼ 96 stars, corresponding to approximately 25% of the total HB population of the cluster (see Extended Data Fig. 6). In addition, seven evolved HB stars are also visible in the CMD at $m_{F275W} < 15.6$, lying exactly on the extension of the extreme-HB sequence and tracing a sort of evolutionary path. According to our adopted BaSTI models[25], these objects have masses below $\sim 0.5$ $M_\odot$. Hence, they likely are *AGB-manqué* stars[47], i.e., objects that completely skip the AGB phase and directly evolve toward the WD stage, because their envelope is so thin that no helium-burning can be ignited in a shell surrounding the carbon-oxygen core after the HB stage. Indeed, this seems to be consistent with the population ratios measured in the two clusters. We adopted appropriate selection boxes to count the AGB and HB populations in the optical and UV-CMDs, respectively (see Extended Data Fig. 6). The AGB-HB ratio turns out to be significantly (at 2 σ level) smaller in M13 ($N_{AGB}/N_{HB} = 30/368 = 0.08 \pm 0.01$) than in M3 ($N_{AGB}/N_{HB} = 45/344 = 0.13 \pm 0.02$), in agreement with the hypothesis that a fraction of HB stars in M13 miss the entire AGB phase (and give rise to 'slow WDs'), while this does not happen in M3, where no HB blue-tail is present.

**Monte Carlo simulations -** To test quantitatively the proposed scenario, we have performed Monte Carlo simulations, using WD cooling models[16] with a thick hydrogen envelope and efficient hydrogen-burning (obtained from the full evolution of progenitors that do not experience the third dredge-up), and their counterpart with hydrogen-burning artificially switched off. The output luminosities have been transformed[11] to the magnitudes of the WFC3 photometric system the same way as done for the standard models[3]. In the relevant magnitude range, the models[16] with no hydrogen-burning have the same cooling times of the "standard" models[3] employed before, despite a hydrogen-envelope a few times thicker.

For the RGB evolution, we have considered BaSTI tracks[25] of appropriate metallicity (Z=0.001), with helium abundance (and evolving mass) in agreement with the distribution derived[24] in each cluster. We emphasise that the evolutionary timescale in the considered RGB portion (see Extended Data Fig. 4) is essentially insensitive to the precise value of the stellar mass around a value of 0.8 $M_\odot$ and to the initial helium abundance, for values up to Y$\sim$ 0.30. The model age has been set to zero at the faint limit of the adopted RGB selection box (see Extended Data Fig 4). For the HB and subsequent WD evolution, we treated M3 and M13 separately.

In M3 we considered the evolutionary timescales of a 0.66 $M_\odot$ BaSTI HB model[25] for the same Z and initial Y, finding $t_{HB} \sim 92$ Myr (this value is essentially independent of the initial Y in the relevant helium mass fraction range). The adopted mass value corresponds to the mode of the HB mass distribution[24]. The total age (RGB plus HB) has then been added to the cooling age of 0.54 $M_\odot$ WD models[16] without nuclear burning. The neglected AGB evolution, between the end of the HB phase and the beginning of WD cooling, provides a negligible contribution to the total age of the WD models.

In the case of M13 we have performed the same kind of simulations, employing the same RGB mass, but two different HB masses for the two WD channels. To simulate the progenitors of slow WDs, we have considered a HB model with mass equal to 0.51 $M_\odot$, which corresponds to the mode of the distribution[24] of the HB masses below 0.56$M_\odot$. The corresponding HB lifetime is ∼ 117 Myr (this value is not affected appreciably by the fact that such a star is predicted to be enhanced in helium by ΔY ∼0.045). For the standard WDs, we have instead considered a HB model with mass equal to 0.60 $M_\odot$ (the mode of the mass distribution[24] of HB stars more massive than 0.56 $M_\odot$), with a lifetime of 93 Myr. Finally, for the evolution of standard WDs we used the same WD models[16], without nuclear burning, employed in M3, whilst for the slow WDs we adopted models[16] with the same mass but with active hydrogen burning. These models have a cooling age of about 140 Myr at the faint magnitude limit considered in our analysis, a factor 1.75 higher than for models without nuclear burning in the envelope. We proceeded as for M3, but separately for the 'standard' and 'slow' WD channels, and then added together the two samples of RGB+HB+WD synthetic stars with the constraint that the number of HB stars from the simulation of the 'slow' WD channel must be equal to 65% of the total number of synthetic HB objects.

**Data Availability:** The data that support the plots within this paper and other findings of this study are available from the corresponding author upon reasonable request.

**Correspondence to:** F.R.Ferraro[1] Correspondence and requests for materials should be addressed to F.R.F. <francesco.ferraro3@unibo.it>

**Acknowledgements:** This research is part of the project COSMIC-LAB at the Physics and Astronomy Department of the Bologna University (see the web page: http://www.cosmic-lab.eu/Cosmic-Lab/Home.html ). The research has been funded by project *Ligh-on-Dark*, granted by the Italian MIUR through contract PRIN-2017K7REXT. JC acknowledges the financial support from the China Scholarship Council (CSC). The research is based on data acquired with the

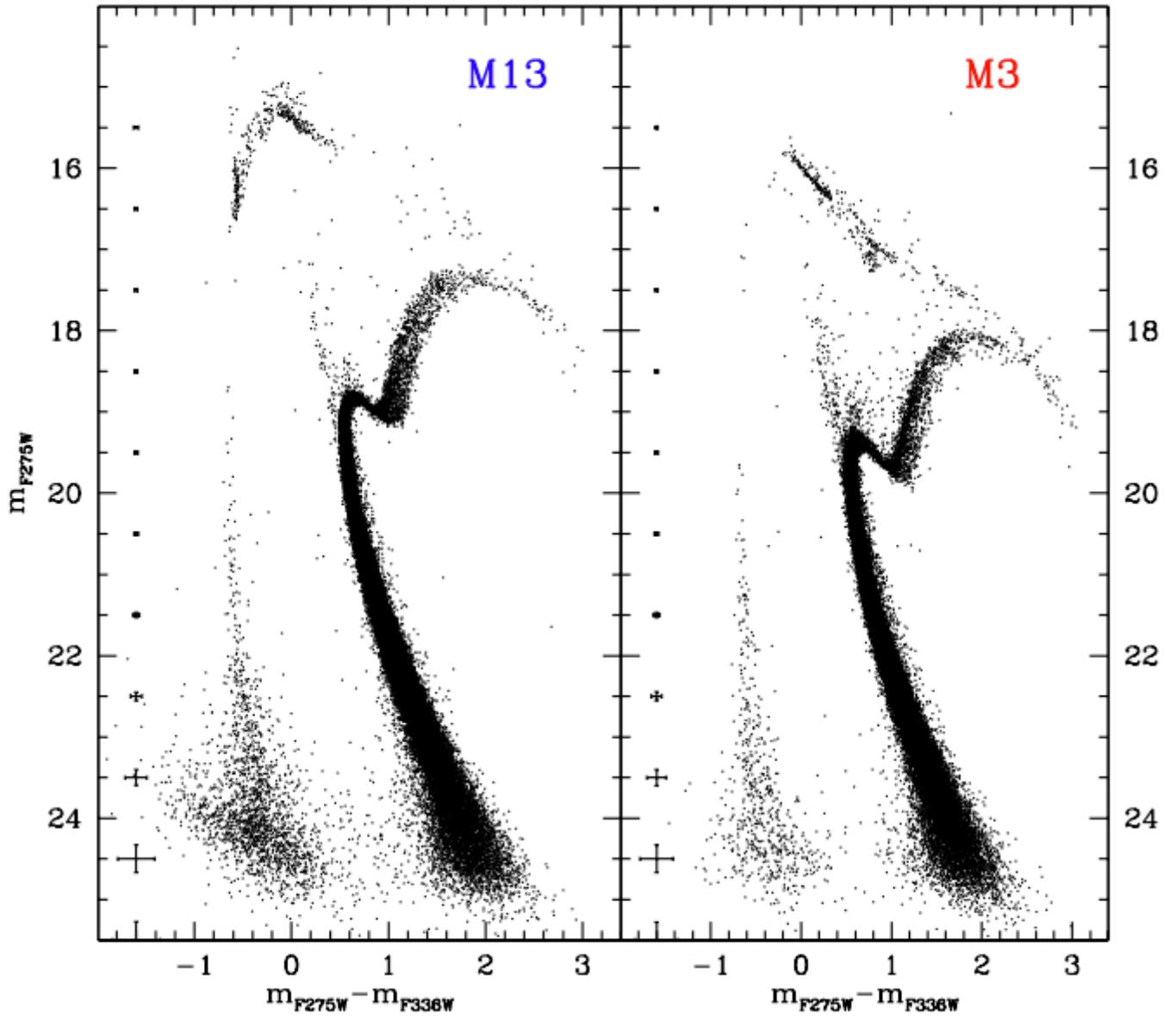

**Figure 1 – The colour-magnitude diagrams of M13 and M3.** Observed ($m_{F275W}$, $m_{F275W} - m_{F336W}$) CMDs of M13 (left-hand panel) and M3 (right-hand panel). The mean photometric errors (1 s.e.m.) in bins of 1 magnitude are also reported on the right side of each panel.

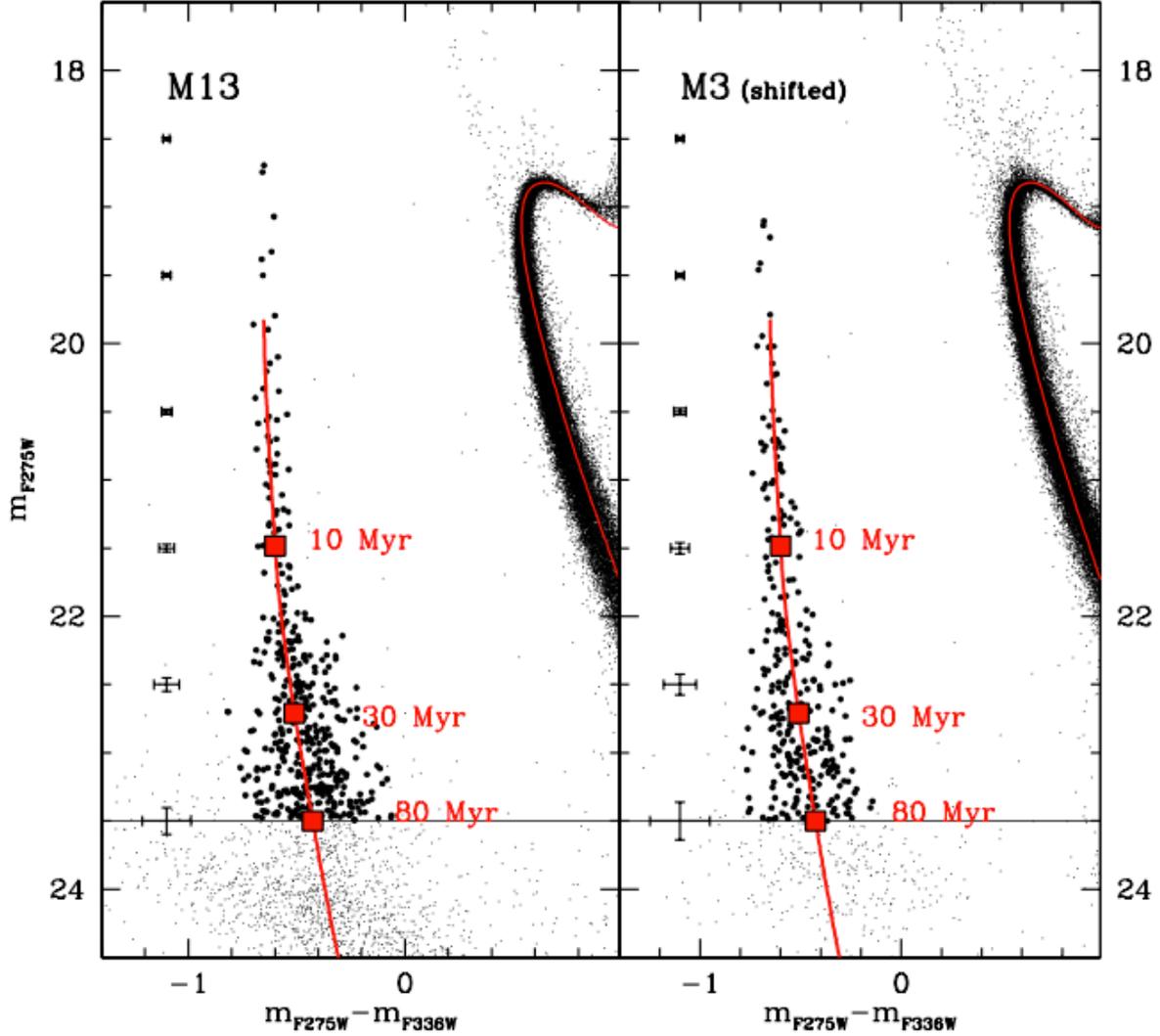

**Figure 2 – The WDs cooling sequences in M3 and M13.** WD cooling sequences in the CMDs of M13 (left) and M3 (right, after the alignment). The WDs selected for the present analysis are marked with large black dots. The red lines are theoretical models reported onto the observational diagram by assuming the distance modulus and the reddening of M13[26], $(m − M)_0 = 14.43$ and $E(B − V) = 0.02$. As apparent, the MS and MS-TO region of both clusters are well reproduced by a 12.5 Gyr old and α−enhanced isochrone from the BaSTI dataset[25], with metal and helium mass fractions $Z = 0.001$ and $Y = 0.246$, respectively, corresponding to a global metallicity [M/H]= −1.27 and an iron abundance [Fe/H]= −1.62. Both the WD cooling sequences are well matched by the cooling track[3] of a 0.54 $M_\odot$ CO-WD with hydrogen atmosphere, transformed[11] to the WFC3 filters. The mean errors (1 s.e.m.) are also marked.

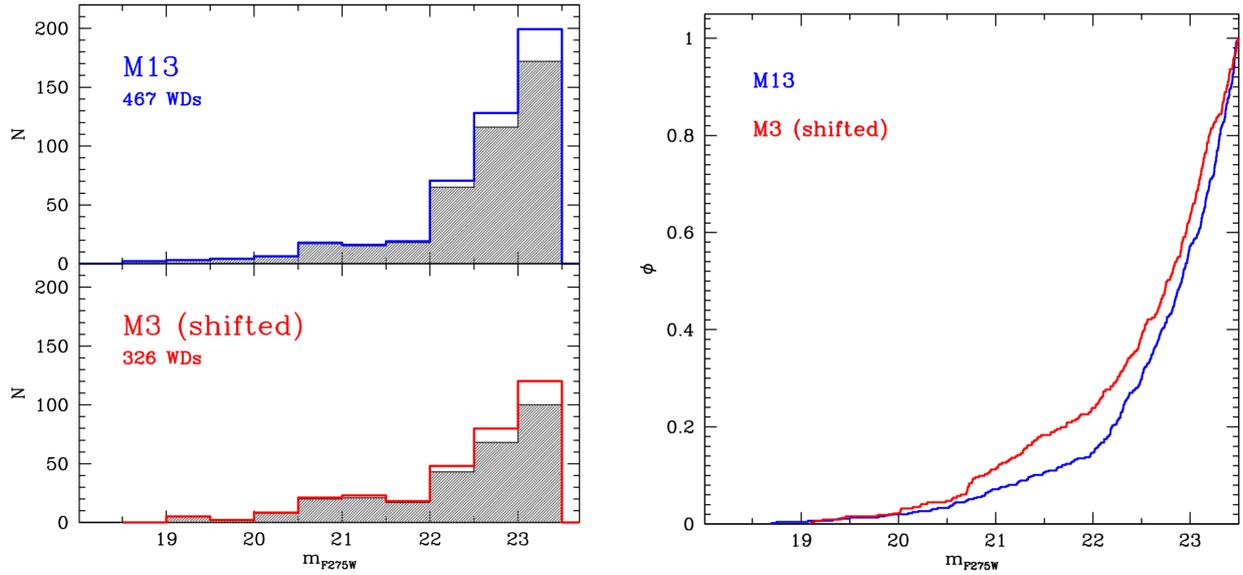

**Figure 3 – Comparing the WD LFs of M3 and M13. a,** Observed and completeness-corrected differential luminosity functions (grey shaded histograms and coloured lines, respectively) of the WDs selected in the two clusters. **b**, Cumulative luminosity functions obtained from the completeness-corrected WD samples in M3 (red line) and M13 (blue line), normalized to the total number of WDs in the respective cluster.

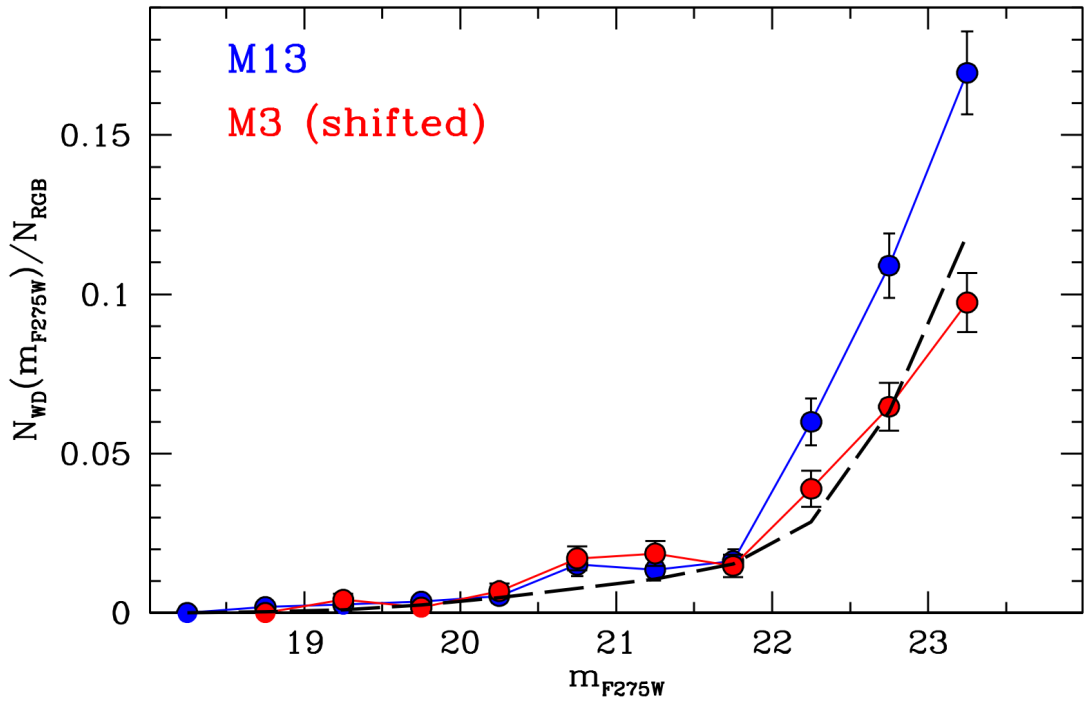

**Figure 4 – The Normalized WD LFs in M3 and M13.** WD differential LFs normalized to the total number of RGB stars selected in the two clusters (see Extended Data Fig.3): red circles for M3 and blue circles for M13. The mean errors (1 s.e.m.) are also reported. The black dashed line corresponds to the theoretical ratio between the standard WD and RGB evolutionary times[3,25] as a function of the WD magnitude.

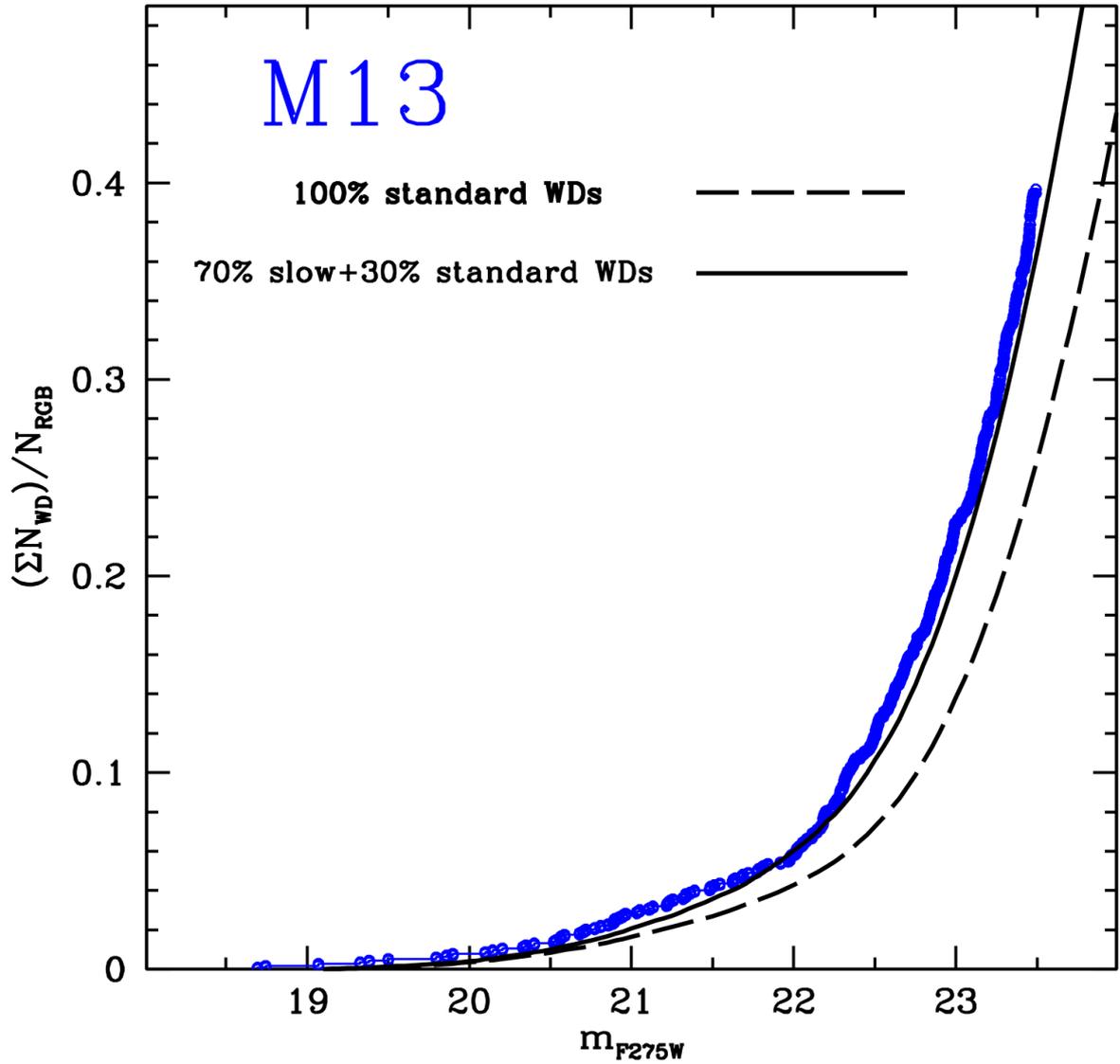

**Figure 5 – Comparison with theoretical cooling models including slow WDs.** Comparison between the cumulative, completeness-corrected, WD luminosity function of M13 normalized to the number of RGB stars (blue line), and the theoretical predictions obtained by assuming WD evolution[16] with no hydrogen-burning (100% standard WDs; dashed line), and a combination of 70% (slow) WDs with active hydrogen-burning and 30% (standard) WDs with no hydrogen-burning (solid line).

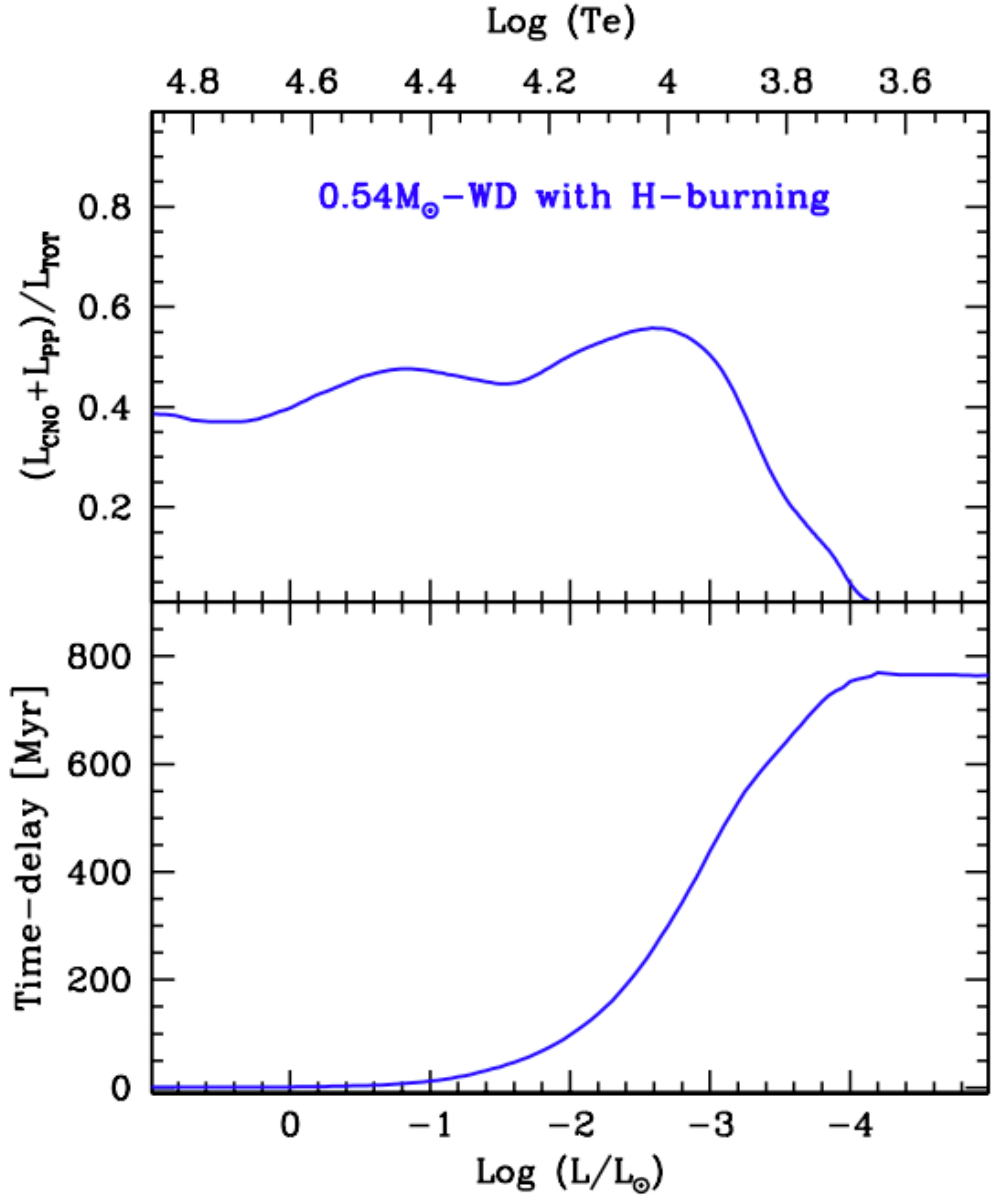

**Extended Data Fig. 1 – The effect of stable H-burning on a low mass WD. a**, Contribution of stable H-burning[15,16] (via PP and CNO chain) to the global luminosity of a low metallicity (Z=0.001), low mass (0.54 $M_\odot$) WD as a function of its decreasing luminosity. H-burning provides a significant contribution (larger than 40%) to the WD luminosity in the brightest portion of the cooling sequence, becoming negligible at $\log(L/L_\odot) \approx -4$ and $\log(Te) \approx 3.7$ (see the temperature scale in the top axis). **b**, Delay[15,16] in the cooling time induced by stable H-burning, with respect to a model without burning. The time delay keeps increasing during the phase of active H-burning and reaches a value as large as ~760 Myr, which then remains constant during the entire subsequent evolution.

| Parameter | M3 (NGC 5272) | M13 (NGC 6205) | Reference |
|---|---|---|---|
| [Fe/H] | −1.50 | −1.53 | Harris (1996)(2010 edition) |
| Age | 12.6 | 12.9 | Denissenkov et al. (2017) |
| $M_V$ | −8.88 | −8.55 | Harris (1996)(2010 edition) |
| $\log(\nu_o)$ | 3.57 | 3.55 | Harris (1996)(2010 edition) |
| $\log(t_{rc})$ | 8.31 | 8.51 | Harris (1996)(2010 edition) |

NOTE—From top to bottom, the listed parameters are: metallicity, age, $V$-band absolute integrated magnitude, logarithm of the central luminosity density (in units of $L_\odot \mathrm{pc}^{-3}$), logarithm of the central relaxation time (in years).

**Extended Data Fig. 2 – Physical parameters of M3 and M13**

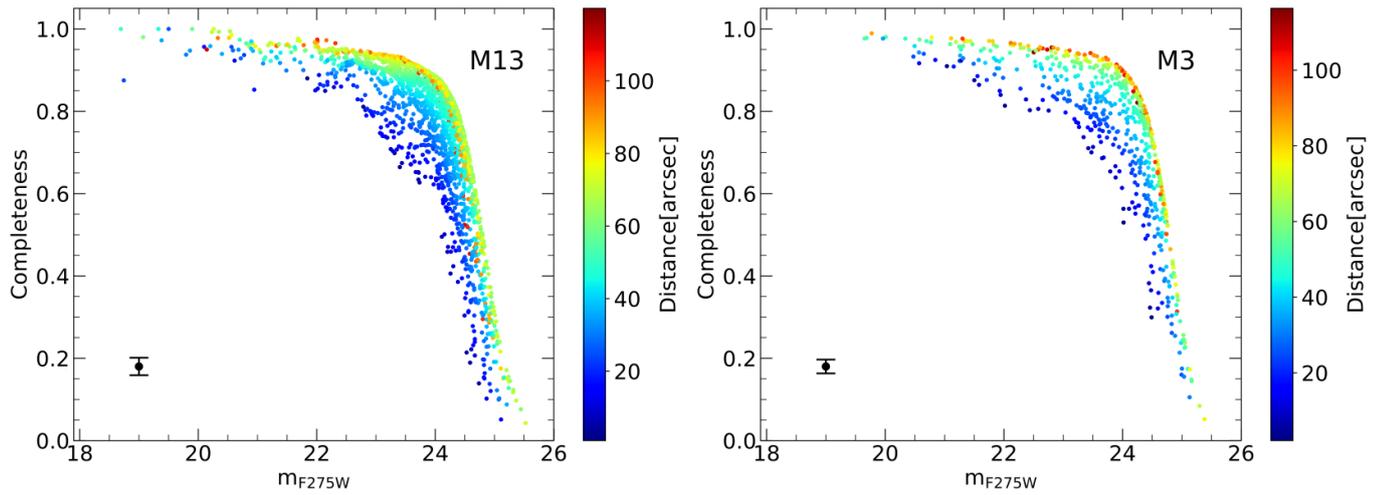

**Extended Data Fig. 3 – The completeness distribution of the WD populations of M13 and M3.** Completeness parameter as a function of the F275W magnitude and colour-coded in terms of the distance from the cluster centre (see colour bars) for each WD detected in M13 (left-hand panel) and in M3 (right-hand panel). The mean error (1 s.e.m.) is also reported.

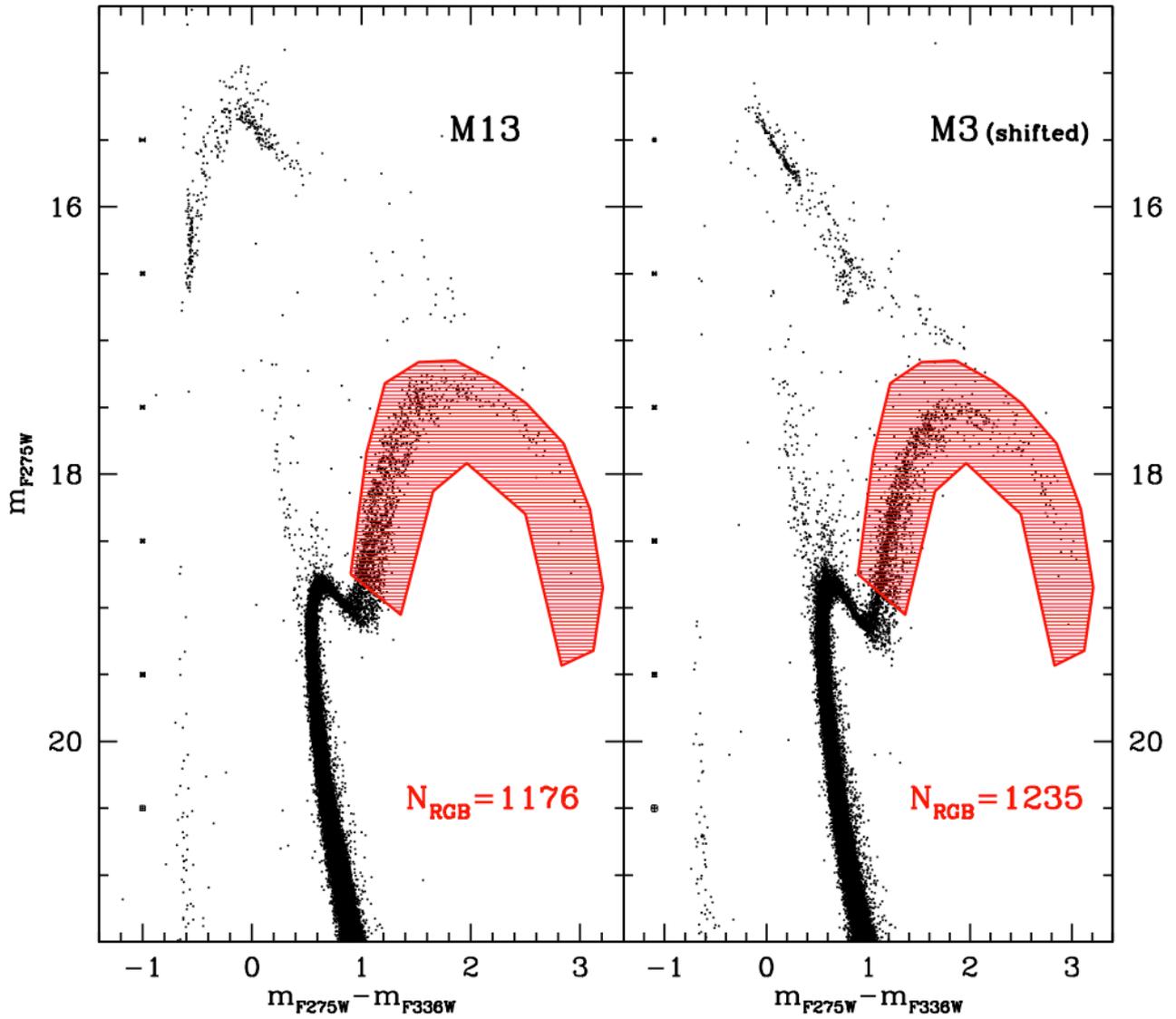

**Extended Data Fig. 4 – The RGB reference population.** Selection box (red shaded area) adopted to define the RGB "reference population" in the observed and realigned CMDs of M13 and M3. The number of red giants counted in each cluster is also marked. The mean errors (1 s.e.m.) are also marked.

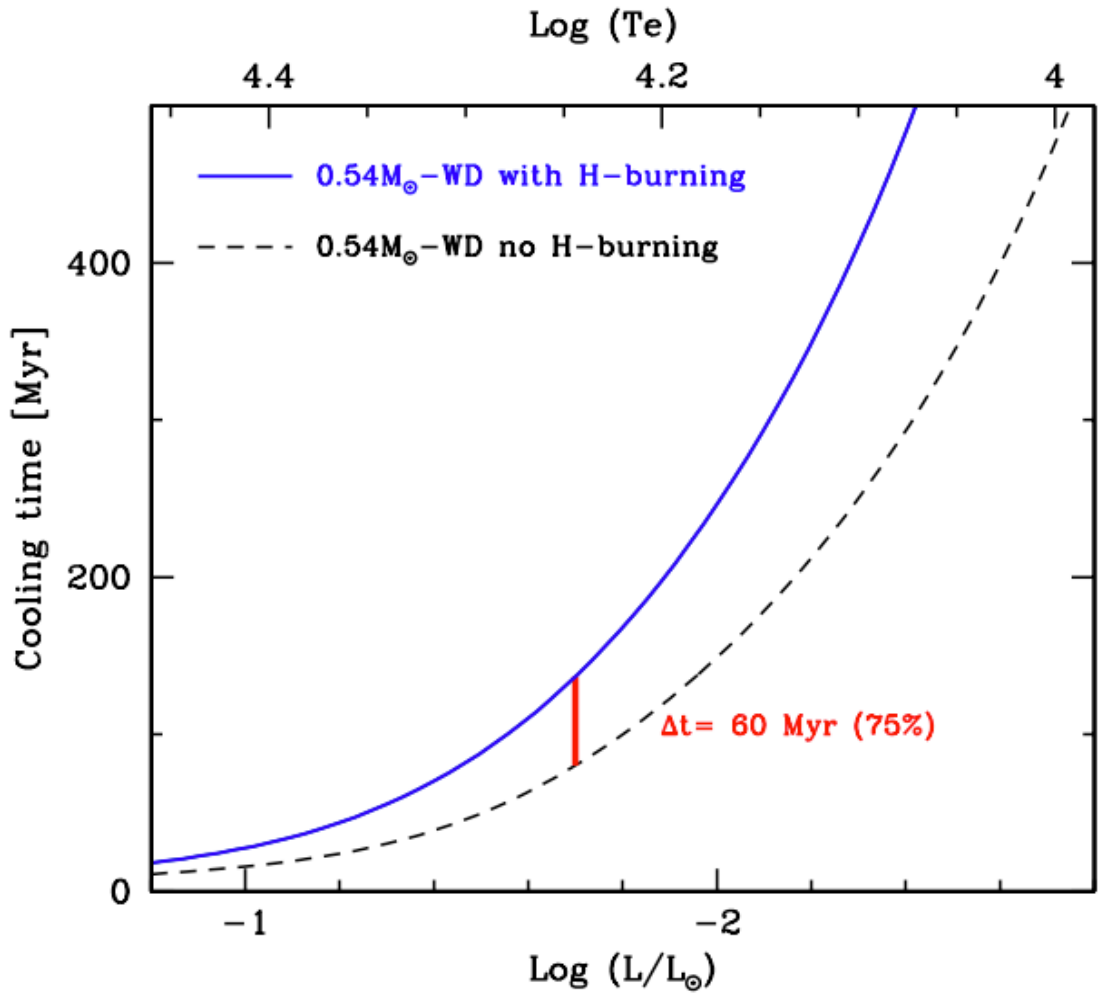

**Extended Data Fig. 5 – WD cooling time for models with and without hydrogen burning.** Comparison between the cooling times of a low metallicity, 0.54 $M_\odot$ WD with and without hydrogen-burning[16] (solid and dashed lines, respectively). The red segment marks the difference in the cooling time at the luminosity of the faintest WD considered in this study, $\log(L/L_\odot)= -1.7$ and reports the absolute difference between the two cooling time values (60 Myr), corresponding to a 75% increase if hydrogen-burning is active, with respect to the 'standard' (no-burning) case.

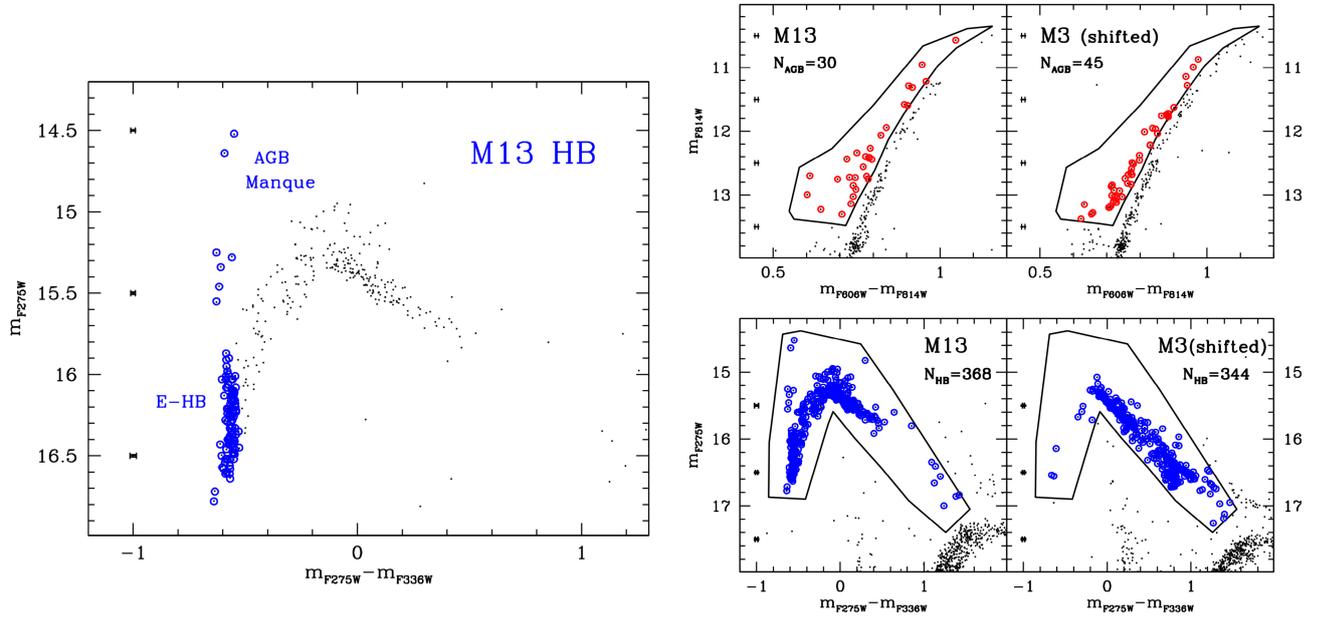

**Extended Data Fig. 6 – HB and AGB populations in M3 and M13. a,** UV-CMD of M13 zoomed in the HB region. The extreme-HB (E-HB) and the 7 candidate AGB-manqué stars are highlighted as blue circles. **b,** AGB and HB selection boxes in the optical- and UV-CMD (top and bottom panels, respectively) for the two clusters. The population star counts are also marked in each panel. The mean photometric errors (1 s.e.m.) are also marked in all panels.